\documentstyle{mn}
\newcommand{\ale}{\ \raisebox{-.3ex}{$\stackrel{<}{\scriptstyle \sim}$}\ }
\newcommand{\age}{\ \raisebox{-.3ex}{$\stackrel{>}{\scriptstyle \sim}$}\ }
\newcommand{\omb}{\Omega_{\rm orb}}
\newcommand{\vsc} {$d\,\Omega_{\rm orb}$}
\newcommand{\lsc} {$d$}
\newcommand{\tsc} {$\Omega_{\rm orb}^{-1}$}

\input psfig

\title[Tilted accretion discs in cataclysmic variables]{Tilted accretion
       discs in cataclysmic variables: \\ Tidal instabilities and
       superhumps}
	
\author[J.R. Murray \& P.J. Armitage]{J.R. Murray$^1$ and P.J. Armitage$^2$\\
	$^1$ The Astrophysical Theory Centre, 
	Australian National University, ACT 0200, Australia \\
 	$^2$ Canadian Institute for Theoretical Astrophysics, McLennan Labs,
	60 St George St, Toronto, M5S 3H8, Canada}	
 	
\begin{document}

\maketitle

\begin{abstract} 
We investigate the growth of tidal instabilities in accretion discs
in a binary star potential, using three dimensional numerical simulations.
As expected from analytic work, the disc is prone to an eccentric
instability provided that it is large enough to extend to the
3:1 resonance. The eccentric disc leads to positive superhumps in 
the light curve. It has been proposed that negative superhumps might 
arise from a tilted disc, but we find no evidence that the companion
gravitational tilt instability can grow fast enough in a fluid disc to
create a measurable inclination. The origin of negative superhumps in the 
light curves of cataclysmic variables remains a puzzle.
\end{abstract}


\section{Introduction}
Classical superhumps are large amplitude variations in the optical light 
curves of close binary systems. The superhump period is typically
 a few per cent longer than the 
binary orbital period. The most plausible explanation for these modulations 
is that resonance of gas orbits in the outer disc with the tidal potential of
the secondary star causes the disc to become eccentric (Whitehurst
1988; Hirose \& Osaki 1990; Lubow 1991a). Due to a combination
of the $m=0$  mode of the tidal potential, 
and pressure and viscous effects, the 
eccentric disc precesses slowly in the prograde direction. Light from the disc
is then modulated on the period of the disc motion relative to the binary 
potential (Murray 1996, 1998).

In this paper we investigate the origin of a second, more puzzling, 
modulation in the light curves of some cataclysmic variables.
These  `negative superhumps' have  a period a few per cent 
{\bf less} than the binary period. Their origin is presently unknown, but 
Patterson et al. (1993) proposed that this was the signature of the 
precessional motion of a {\bf tilted disc}. This is an attractive 
explanation of the observations, but if correct it poses the difficult
question of how such a tilted disc can arise in cataclysmic variable (CV)
systems. 

In this paper we investigate various methods of exciting a tilt or warp in
a CV disc. The next section is a  summary of the observations of the new
periodicity.  In section 3 we discuss the conditions under which tidal
instability might induce a disc tilt, which we investigate numerically 
in sections 4 and 5. Section 6 discusses the differences between CVs and X-ray
binaries such as Her X-1, where the existence of warped discs is 
well established. Section 7 summarises our results, and discusses
what observations are required to further investigate the
possible existence of tilted discs in CVs.

\section{Observations of Negative Superhumps}

An optical modulation with a period slightly
shorter than the orbital period, $P_{\rm orb}$, has been seen in several
cataclysmic 
variables (Patterson et al. 1993; Patterson
et al. 1997). Here we briefly describe a few systems that exhibit these
so-called `negative superhumps'.

The nova V603 Aquilae has a high mass transfer rate which implies a
hot accretion disc (Patterson \& Richman 1991). These systems are likened 
to dwarf novae in permanent outburst, and by analogy with those systems are
expected to have a high ($\sim 0.1$) $\alpha$ viscosity 
parameter (Shakura \& Sunyaev 1973). V603 AqI's orbital period
($P_{\rm orb}=199.0$ minutes) is longward of the period gap but still short
enough to suggest that the disc could maintain contact with the $3:1$
eccentric Lindblad resonance and become significantly eccentric.
Indeed V603 AqI consistently exhibits  a photometric signal with period 
$P_{\rm sh}^+ \simeq 210$ minutes that is thought to be the signature
of an eccentric, precessing disc (Patterson \& Richman 1991; Patterson
et al. 1997). Hence V603 AqI is sometimes referred
to as a {\bf permanent superhumper}. 

Patterson et al. (1997) reported
the appearance  of a  second  signal, with a period 
$P_{\rm sh}^- = 193.0$ minutes, in the V603 AqI light curve. This
`negative superhump' coexisted with and became brighter than its
better known counterpart. They found the positive and
negative superhump periods varied on similar time scales and were
negatively correlated. In other words the negative superhump period
was shortest when the positive superhump was longest. Patterson et
al. (1997) hypothesised that the accretion disc was simultaneously eccentric,
and tilted. The prograde precession of the disc's semi-major axis (as
viewed in the inertial frame) gave
rise to the positive superhump signal, whilst the retrograde
precession of the disc's line of nodes was responsible for the
shorter, negative superhump periodicity in the light curve.

The SU~UMa type dwarf novae V503 Cyg (Harvey et al. 1995)
and V1159 Orionis (Patterson et al. 1995) 
 have both displayed large amplitude
negative superhumps during  quiescence and  normal outburst.
In the case of V503 Cyg the normal, positive superhumps were only seen
during supermaximum, whereas in V1159 Orionis they persisted well 
beyond superoutburst and into the next normal outburst.
Negative superhumps are not limited to short period systems
however. For example
TV~Col, with $P_{\rm orb} = 5.5$ hours, displays a photometric period 
$P_{\rm sh}^- = 5.2$ hours (Hellier 1993).

\section{The tidal inclination instability}
A fluid disc in a binary potential is subject to both eccentric and
tilt instabilites at the 3:1 resonance (Lubow 1992a). Since it appears
well-established that {\em positive} superhumps are the signature of 
an eccentric, precessing disc, we first consider whether the companion
tilt instability can give rise to a negative superhump signal. In 
Lubow's analysis (1992a), the growth rates of these instabilities are (for
the idealised case of a narrow gaseous ring at the resonance),
\begin{eqnarray}
 \lambda_{\rm ecc} & = & 2.08 \ \omb\ q^2 \,{ r_{\rm res} \over W }
  \label {eqn:gr1}\\
 \lambda_{\rm inc} & = & 0.0398 \ \omb\ q^2\,{ r_{\rm res} \over W }, 
\label{growth_rate}
\end{eqnarray} 
where $W$ is the width of the resonance at radius $r_{\rm res}$, 
$q$ is the mass ratio ($M_{\rm secondary} / M_{\rm primary}$),  
and $\omb$ is the binary angular velocity. To
develop a tilt via this instability, we require that the growth time 
of the tilt $\tau_{\rm inc} \approx 1 / \lambda_{\rm inc}$ must be
much less than the drift time for material through the resonance,
$\tau_{\rm adv} = W / \vert v_{r} \vert$. Making use of 
$v_{r} = - 3 \nu / 2 r$, and assuming an $\alpha$ prescription 
for the viscosity, $\nu = \alpha c_{\rm s}^2 / \Omega_K$, we obtain,
\begin{equation}
 \alpha \ll {2 \over 225} \ q^2 \left( H \over r \right)^{-2}
\label{alpha_limit}
\end{equation}
where $H = c_{\rm s} / \Omega_{\rm K}$ is the disc scale height,
$c_{\rm s}$ is the sound
speed, and $\Omega_{\rm K}$ is the Keplerian angular velocity. For $q = 0.25$ 
and $H / r = 1 / 20$, the constraint on $\alpha$ from this
analysis is $\alpha \ll 0.2$. Given that typical estimates 
for $\alpha$ in high-state dwarf novae are $\age 0.1$
(Cannizzo 1993), this immediately suggests that exciting a
tilt via the instability is difficult. However it is also clear
that these estimates are rather crude, and so in the next section
we use numerical simulations to explore whether a gravitational
tilt instability can be excited in the conditions of CV discs.

\section{Numerical method}

\subsection{SPH code}

A smoothed particle hydrodynamics (SPH) code designed specifically for
accretion disc problems has been described in detail in 
Murray (1996) and Murray (1998).  
Modifications have been made to extend the code to three dimensions,
and to include adaptive resolution (i.e. to allow for variable
smoothing length $h$). For disc simulations, the standard SPH linear 
artificial viscosity
term (as described in Monaghan 1992) can be modified slightly in order
to improve its effectiveness as a source of shear viscosity in the SPH
equations of motion. The modifications are described in some detail in
Murray (1996) and Murray (1998). By taking the SPH equations to the
continuum limit, it can be shown that this artificial viscosity term
generates a shear viscosity

\begin{equation}
\nu = \kappa\,\zeta\,c_{\rm s}\,h,
\label{eqn:SPHvic}
\end{equation}
where  $h$ is the SPH smoothing length, and $\zeta$ is
the (dimensionless) artificial viscosity parameter. For the cubic
spline in three dimensions, $\kappa$ is analytically found to be
$1/10$.

In the simulations shown in Murray (1996) and Murray (1998) the smoothing
length had a unique fixed value. In the calculation described
here we allow $h$ to vary with both space and time. When 
evaluating the pressure-viscosity interaction between two particles,  $i$
and $j$, we choose to symmetrise with respect to the smoothing length
and simply use  $\overline h = \frac{1}{2}\, (h_i+h_j)$ in place of $h$ in the
momentum equation. The density is evaluated using the form that
appears as equation (6.3) in Monaghan (1992), namely
\begin{equation}
\rho_i=\sum_j m_j\,W({\bf r_i}-{\bf r_j},h_j).
\end{equation}
We use 
\begin{equation}
h_i=K\,\rho_i^{-1/3}
\end{equation}
to alter particle smoothing lengths at every time step. Here K is a
constant set at the start of the calculation to ensure that each
particle $i$ has a reasonable number of neighbours, $N_i$. 
Neighbour numbers were
monitored closely throughout the calculation and averaged near 75.
The variable $h$ implementation of the code was tested against
a fixed $h$ version.

\subsection{Disc tilt and eccentricity}
\begin{figure*}
\mbox{\psfig{figure=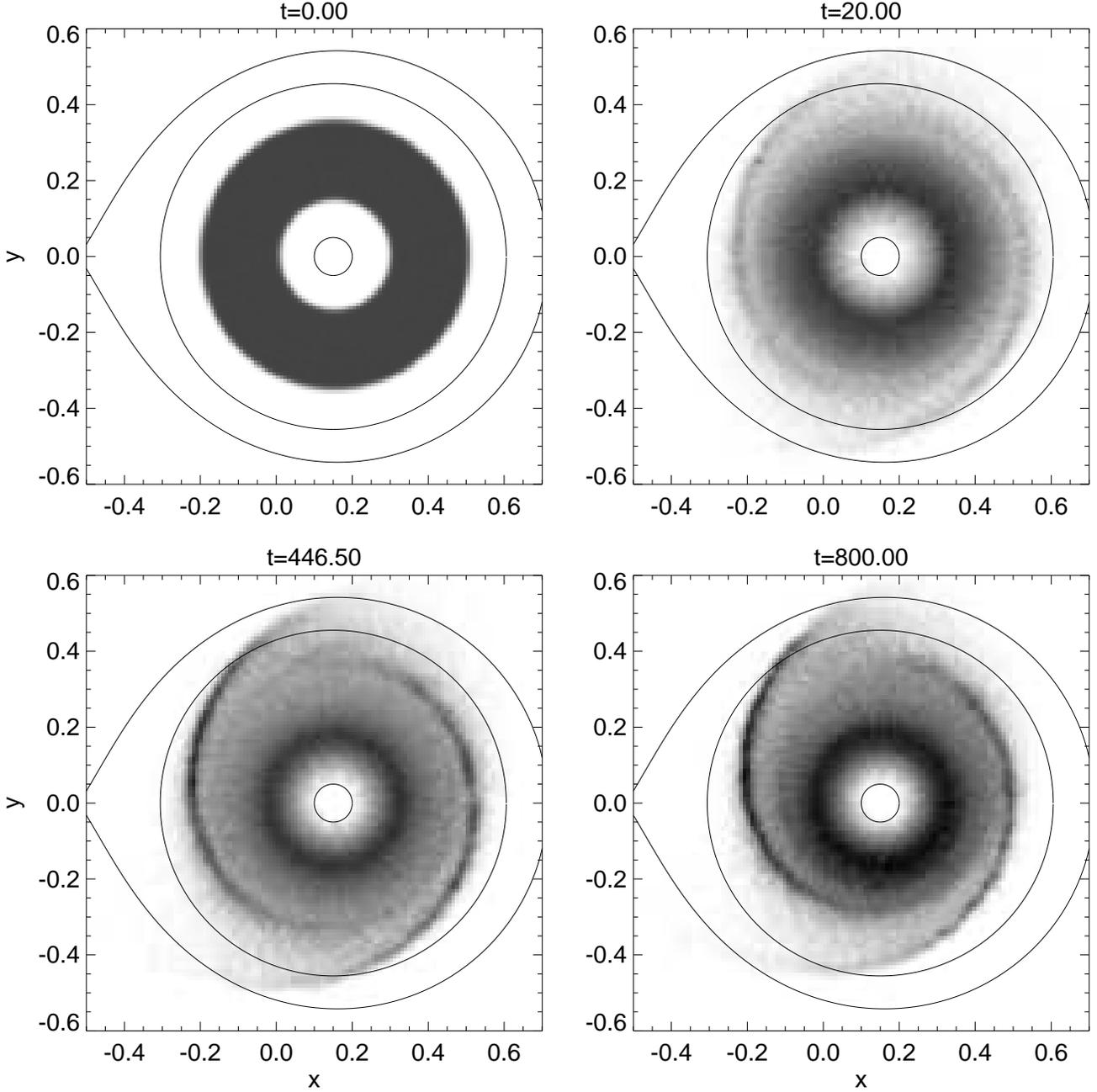,width=18cm}}
\caption{Grey scale column density maps of the disc at the times
shown. The same density scale is used in all frames. The Roche lobe of
the primary, the inner boundary at $r_{\rm wd}$, and the approximate
location of the tilt resonance are marked as solid lines.}
\label{fig:discthick}
\end{figure*}

We follow the evolution of tilt and eccentricity in the disc
simulations by decomposing the particle distribution into Fourier
components with argument $(l\theta - m\,\omb t)$, where $l$ and
$m$ are integers, and $\theta$ is azimuth in the inertial frame. 
As in Lubow (1991b) we define  the 
strength of the $(l,m)$ mode 
\begin{eqnarray}
S_{l,m}(t)=({S_{\cos,\cos,l,m}(t)}^2&+&{S_{\cos,\sin,l,m}(t)}^2
\nonumber\\
+{S_{\sin,\sin,l,m}(t)}^2&+&{S_{\sin,\cos,l,m}(t)}^2)^\frac{1}{2}.
\label{eqn:mdstrength}
\end{eqnarray}
The disc
tilt is the $(1,0)$ Fourier component of the vertical displacement
of the particles.
The component modes of the vertical displacement are defined as follows:
\begin{eqnarray}
S^z_{\sin,\cos,l,m}(t)&=&\frac{2\,\omb}{\pi 
N (1+\delta_{l,0})(1+\delta_{m,0})}
\nonumber\\
\times \int_t^{t+2\pi\omb^{-1}} &\sum_{p=1}^N&z_p(t)\sin(l\theta_p) 
\cos(m\omb t')\,\rm{dt'},
\label{eqn:zstrength}
\end{eqnarray}
where $\theta_p$ is the angular position of particle $p$ and $N$ is
the total number of particles. $\delta$ is the
Kronecker delta. We use a superscript $z$ to distinguish modes of the
vertical displacement from planar modes.

As in Murray (1996) and Murray (1998), we use the interstellar
separation $d$, the binary mass $M$, and the inverse of the orbital
angular velocity \tsc, as the length, mass and time scalings for our
simulations.  

\section{Results}
\begin{table}
\caption{Summary of parameter settings used in the tilt calculation}
\label{tab:results}
\begin{tabular}{@{}lccccccc}
Parameter & $c_{\rm s}$ & $q$ & $\zeta$ & $h_{\rm max}$ & $r_{\rm wd}$ & $\Delta t$\\
Units&  \vsc & & & \lsc  & \lsc & \tsc \\
Value & 0.05 & 3/17 & 1.0 & 0.02 & 0.05 & 0.01\\
\end{tabular}
\end{table}

The tidal warping of accretion discs was studied by Larwood et
al. (1996). They used three-dimensional SPH simulations to investigate
the warping, precession and truncation of a disc {\em that initially
had a non-zero inclination to the binary plane}. In the case of CVs we
must determine whether a tilt can be excited in a disc that
is initially co-planar with the binary. In the simulation described
here we investigate whether such a tilt can be generated by the
inclination instability of Lubow (1992a).

\subsection{Initial conditions}

The initial condition for this simulation was an annulus  composed of
21 concentric rings of particles spaced $\Delta r = 0.01\,d$
apart. Each ring consisted of 9 layers of particles centred on the
midplane, spaced $\Delta z = 0.01\,d$ apart. The disc then initially
contained $29592$ particles. Thereafter mass was added to the disc
midplane at the circularisation radius $r_{\rm circ} = 0.1781\,d$ 
at the the rate of one particle per time step $\Delta t = 0.01$ \tsc.

We assumed the disc to be isothermal, with the sound speed $c_{\rm s}=0.05$
\vsc. A binary mass ratio $q=3/17$ was used. In previously described
simulations (Murray 1998) with this value $q$, the disc rapidly
became eccentric. We set the maximum value for particle smoothing
lengths to be $h_{\rm max}=0.02\,d$. We used an open inner boundary at
$r_{\rm wd}=0.05\,d$. Particles ending a time step with $r<r_{\rm wd}$
were considered to have been accreted by the white dwarf and were
removed from the simulation. Particles that either returned to the
secondary or were flung to very large radii ($r > 0.9\,d$) were also 
removed from
the calculation. The parameter settings used in the simulation are
summarised in table~\ref{tab:results}. The calculation ran for $800$
\tsc\,(approximately 127 orbits of the binary).

\subsection{Disc structure}
\begin{figure}
\mbox{\psfig{figure=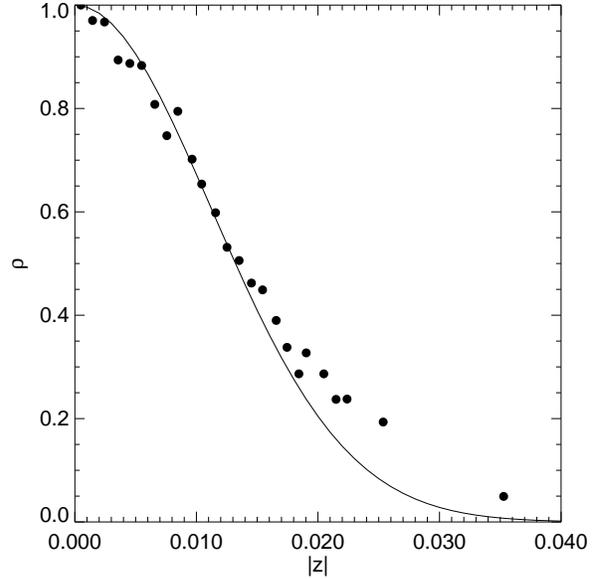,width=8cm}}
\caption{Mean density (scaled to the midplane density) as a function of
height above the midplane for particles in an
annulus $0.348\,d < r < 0.352\,d$ (filled points) for the simulation
at time $t=446.50$ \tsc. At this time there are 34,406 particles in the disc. 
The appropriate isothermal atmosphere
($\rho(z)=\rho(0)\,\exp(-z^2/2H^2)$ where $H=rc_{\rm s}/v_\phi$ is the density scale height) is shown as a solid line.}
\label{fig:rhoz}
\end{figure}

\begin{figure}
\mbox{\psfig{figure=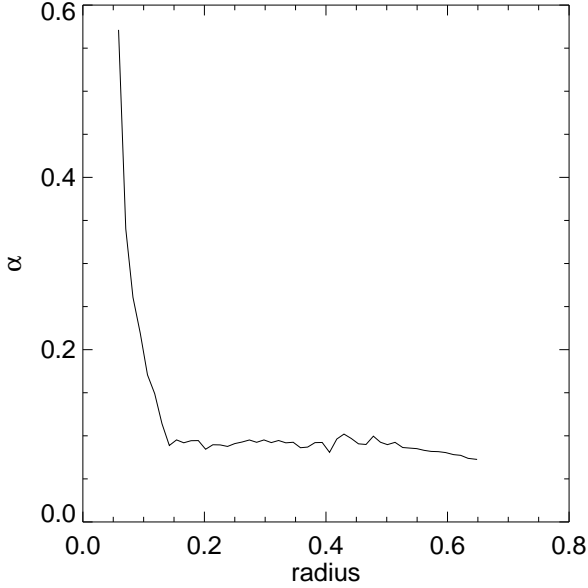,width=8cm}}
\caption{Estimated value of the Shakura-Sunyaev viscosity parameter
$\alpha$ as a function of radius for the simulation at time $t=446.50$
\tsc. At this time the disc contained 34405 particles, and the mean
number of neighbours was $77.5$.}
\label{fig:alpha}
\end{figure}

\begin{figure}
\mbox{\psfig{figure=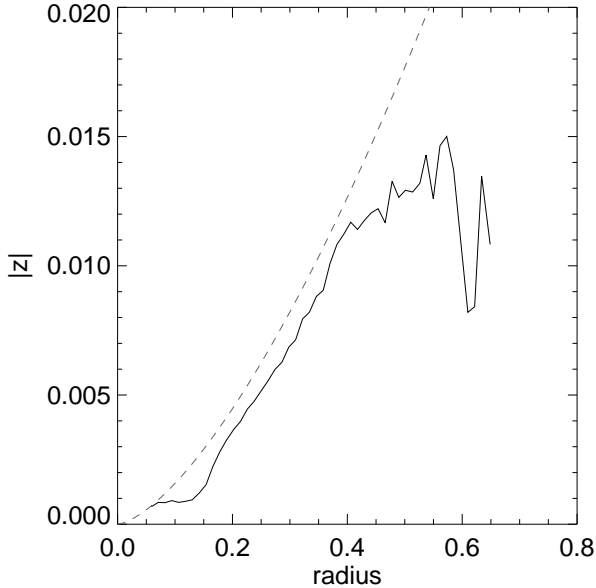,width=8cm}}
\caption{Average height of particles above the midplane 
$\alpha$ as a function of radius for the simulation at time $t=446.50$
\tsc. The scale height for an isothermal disc with $c_{\rm s}=0.05$ \vsc\, is
plotted as a dashed line.}
\label{fig:height}
\end{figure}

Figures~\ref{fig:discthick} through~\ref{fig:height} summarise the 
structure of the model disc. 
Fig.~\ref{fig:discthick} is a sequence of surface density maps showing
the initial condition; the short
time scale
response; the later evolution; and the
final state of the calculation. As expected for this mass ratio, the
disc rapidly expanded to reach the Roche lobe of the primary, and to
overlap the $3:1$ eccentric resonance.
Although there is considerable strength in the eccentric mode by the
end of the simulation (see below), it is the $(2,2)$ tidal mode that
is most apparent in each of the frames.
 
Fig.~\ref{fig:rhoz} plots the azimuthally averaged 
vertical density profile for 
the annulus $0.348\,d<r<0.352\,d$, for comparison against the
density profile of the corresponding isothermal atmosphere. With 
this particle number and choice of disc sound speed we obtain 
acceptable resolution of the disc's vertical structure over 
almost an order of magnitude in density. Vertical velocities were
subsonic everywhere in the disc, further indicating that the vertical
resolution is satisfactory. For a cooler disc, 
with $c_{\rm s}=0.02$ \vsc\, we were unable to adequately resolve 
the vertical structure.

The radial structure of the disc in the latter stages
of the simulation is presented in Figures~\ref{fig:alpha}
and~\ref{fig:height}. 
From Fig.~\ref{fig:alpha} it is apparent that particle resolution is
inadequate at small radii $r \ale 0.15\,d$.
Exterior to this radius, the effective Shakura-Sunyaev $\alpha$ 
parameter,
\begin{equation}
\alpha\simeq \frac{1}{10}\,\zeta h \,\frac {\Omega}{c_{\rm s}},
\label{eq:shakest}
\end{equation} 
is approximately constant at $\alpha=0.1$.
We expect this to be a reasonable value for alpha for 
nova-like systems such as V603 Aquilae, which have shown 
negative superhumps. It might be too high (and, the disc 
too hot) for dwarf novae in quiescence. 
In Fig.~\ref{fig:height} the mean height of particles above the disc
plane is shown as a function of $r$. Out to a radius $r \simeq
0.4\,d$, this profile is consistent with the $r^{3/2}$ scaling of an
isothermal axisymmetric disc. At larger radii however, the influence
of the companion and the non-axisymmetry of the disc breaks the
$r^{3/2}$ scaling relation.
\subsection{Mode strengths}

\begin{figure}
\mbox{\psfig{figure=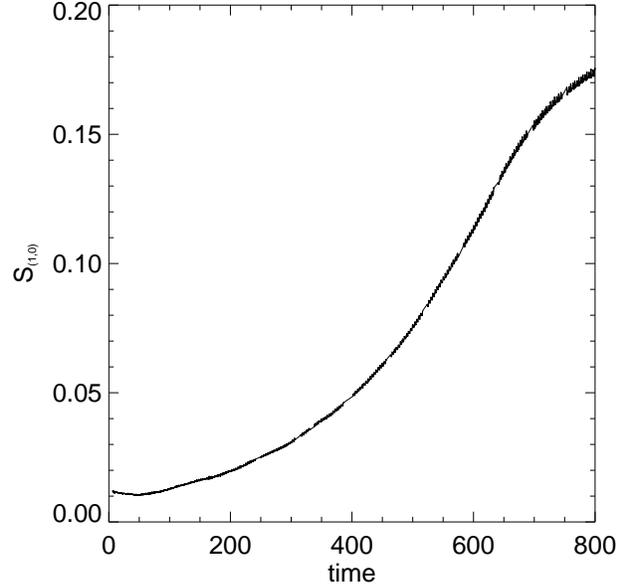,width=8cm}}
\caption{Eccentric mode strength as a function of time. }
\label{fig:eccyd}
\end{figure}

\begin{figure}
\mbox{\psfig{figure=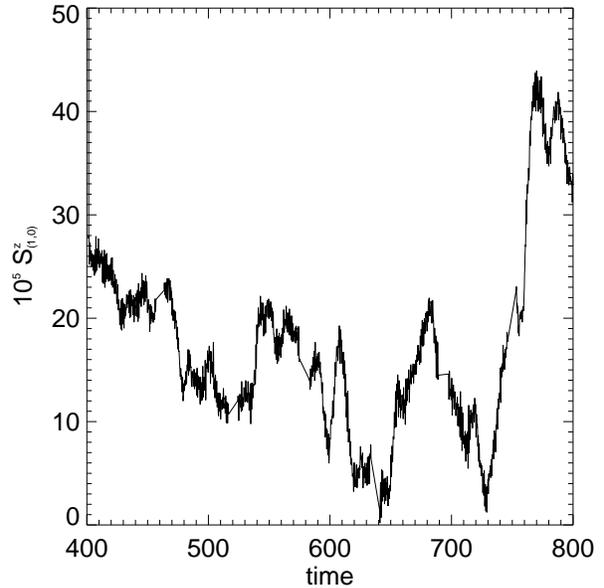,width=8cm}}
\caption{Strength of the (1,0) mode of the 
vertical displacement as a function of time. 
For graphical convenience we have actually plotted 
$10^5 \,\times\,S^z_{(1,0)}$. The strength of this mode was only
calculated for $t > 400$ \tsc.}
\label{fig:eccyz}
\end{figure}

\begin{figure}
\mbox{\psfig{figure=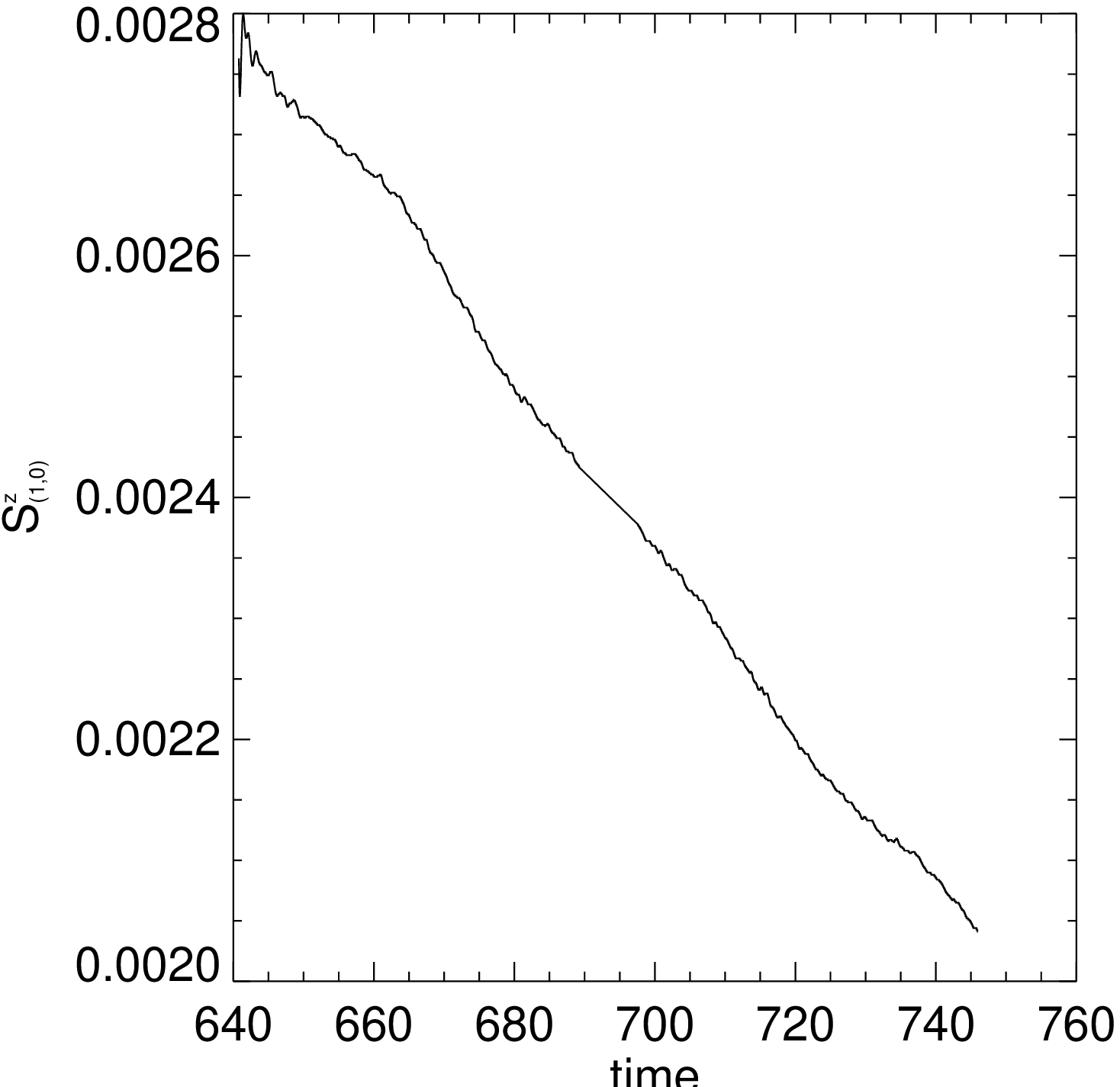,width=8cm}}
\caption{Strength of the (1,0) mode of the 
vertical displacement as a function of time for disc with perturbation
$z^\prime_i=z_i+0.02\,r_i\,\cos(\theta_i)$ applied at time $t=634.50$ \tsc.}
\label{fig:eccyz2}
\end{figure}

Fig.~\ref{fig:eccyd} shows the eccentric mode strength in the disc, 
defined in equation~(\ref{eqn:mdstrength}), as a function of time. 
For times $t \ale 550$ \tsc\, the eccentric mode grew exponentially, 
with a growth rate of $\lambda=0.0044\pm0.0001 \,\omb$. 
When equation~(\ref{eqn:gr1}) is
corrected to allow for a broad disc (see Lubow 1991a), we obtain the
analytic estimate for the growth rate, $\lambda_{\rm t}\simeq
0.02\,\omb$. The simulation eccentricity therefore grew
significantly more slowly than analysis predicted, and also more slowly
than in previous numerical calculations 
(Murray 1996, 1998). However, the discs described in those papers were colder
and more viscous than the disc detailed here. 
Consequently the tidal modes in those simulations were much
weaker than in this calculation. Lubow (1992b) numerically showed that
the non-resonant tidal response of the disc weakens the eccentric
instability, an effect not accounted for in
equation~(\ref{eqn:gr1}). 
Hence the reduced eccentricity growth rate in this simulation is
consistent with our understanding of the resonance.

In Fig.~\ref{fig:eccyz} we plot the strength of the disc tilt as defined in 
equation~(\ref{eqn:zstrength}). The variations seen in this 
plot are of very small ($\ll 1$ degree) amplitude, and are
consistent with noise in the simulation.
{\em No significant tilt is generated.} We ran a short comparison
calculation, taking the disc output at $t=634.50$ \tsc\, and adding a
one degree inclination to it. Mass addition, and all other parameters
were kept as before.
The subsequent evolution of the tilt
mode strength is shown in Fig.~(\ref{fig:eccyz2}). With the addition
of the tilt, $S_{(1,0)}^z$ is almost two orders of magnitude larger
than at any stage of the original simulation. The mode strength
however then decayed with a time scale $\simeq 400$ \tsc, which agrees well
with the simple estimate $M_{\rm disc}/\dot{M} \simeq 350$ \tsc. This
is as expected if the continued addition of mass simply dilutes any
initial disc inclination.

\section{Comparison with X-ray binaries}

The prototypical example of a warped, precessing accretion disc 
is that of the X-ray binary Hercules X-1. Recently it has been
suggested that this, and similar systems, develop warps as a
result of the radiative warping instability analysed by 
Pringle (1996; see also Maloney, Begelman \& Pringle 1996;
Maloney \& Begelman 1997). Here, we briefly summarise
why the same mechanism does not work for CVs, except 
in a few, exceptional, circumstances.

The radiative warping instability arises because reprocessed
radiation from a central luminous source is re-emitted 
non-axisymmetrically from an optically thick disc possessing
a warp. Growth occurs if the disc is larger than some critical 
radius, $r_{\rm crit}$, given for an isothermal disc (Pringle 
1997; for the general case see Maloney, Begelman \& Nowak 1997)
by,
\begin{equation}
 { r_{\rm crit} \over r_{\rm s} } \simeq 2 \pi^2 \left( \eta^2 \over 
 \epsilon^2 \right),
\label{warp_criteria}
\end{equation}
where $r_{\rm s} = 2 G M / c^2$ is the Schwarzschild radius of the accretor,
$\epsilon = L / \dot{M} c^2$ is the radiative efficiency of the flow,
and $\eta$ is the ratio of the `vertical' viscosity (that acting to 
reduce the relative inclination of disc annuli) to the shear viscosity 
in the plane (see Papaloizou \& Pringle 1983 for a formal definition).
For accretion powered cataclysmic variables, 
\begin{equation}
 \epsilon \approx { {GM} \over {r_{\rm acc} c^2} } \sim 10^{-4},
\label{eps_cv}
\end{equation} 
and the instability criterion, equation (\ref{warp_criteria}), requires
$r_{\rm crit}$ prohibitively large, of order $10^{14} \ {\rm cm}$ -- much 
greater than the actual extent of the disc. Steady CV discs are thus 
generally stable against radiative warping.

Evading this conclusion requires either a higher radiative efficiency
or an appeal to non-steady-state conditions. The former is possible
if there is steady nuclear burning of accreted material on the surface
of the white dwarf. The radiative efficiency in this regime is as high as
$\epsilon \approx 7 \times 10^{-3}$, leading to an estimate for 
$r_{\rm crit}$ of the order of $10^{11} \ {\rm cm}$. With the substantial 
theoretical uncertainties, for example in $\eta$, this could lead to 
warping provided the disc was large enough and optically thick. 
However the requirement for nuclear burning limits these
possibilities to supersoft X-ray sources (Southwell, Livio \& Pringle 1997)
rather than lower accretion rate CVs.

The wide range of transient behaviour seen in CVs suggests a number
of scenarios where non-steady-state warping effects might be important.
For example, the outer cool parts of a dwarf nova disc might be 
susceptible to warping from the radiation emitted by hotter inner 
regions that were decaying from outburst, or from radiation from 
a cooling white dwarf heated during the high state accretion event. 
Unfortunately, while these scenarios produce a disc that is formally
unstable by equation (\ref{warp_criteria}), the time scale for 
growth is generally much greater than the duration of the transient
state. The time scale for growth is (Pringle 1996),
\begin{equation}
 \tau_{\rm rad} = { {12 \pi \Sigma r^3 \Omega_K c} \over L }
\label{tau_rad}
\end{equation}
which for highly optimistic CV parameters is,
\begin{eqnarray}
 \tau_{\rm rad} \simeq 5 \times 10^7  
 \left( { M \over {0.6 \ M_\odot} } \right)^{1/2}
 \left( { \Sigma \over {10 \ {\rm g\, cm^{-2}}} } \right) \\ \nonumber
 \times \left( { r \over {10^{10} \ {\rm cm}} } \right)^{3/2}
 \left( { L \over {2 \times 10^{33} \ {\rm erg\, s^{-1}}} } \right)^{-1} \ {\rm s}.
\label{tau_CV}
\end{eqnarray} 
We therefore conclude that there is insufficent luminosity to generate
a finite warp in this manner. Moreover there is no sign that the 
systems with observed negative superhump signals {\em are} universally prone 
to dramatic transient changes in accretion rate, as would certainly 
be required for this mechanism to operate.

\section{DISCUSSION}
We have found that the tidal inclination instability is too weak to 
generate a significant tilt in an accretion disc in the high state. 
However, neither our time scale analysis, nor our simulations can 
rule out such instability occurring in short period systems with quiescent
discs. In a disc of very weakly interacting particles, vertical 
instability as opposed to tilt instability was relatively easy to 
excite at the $3:1$ resonance.
Thus, although it seems unlikely that the inclination instability
discussed by Lubow (1992a) can account for the negative superhump
phenomenon in currently observed systems, the instability might be 
important in low state dwarf nova, particularly those such as
WZ Sge systems where the quiescent $\alpha$ parameter is thought 
to be very low (though see also Warner, Livio \& Tout 1997 for a 
contrary view). Although the contribution to the observed emission
from the disc is small in these systems, they are likely to be
the best places to investigate tidally-driven disc tilts.

Although this work does not categorically rule out the possibility of
tilted discs in cataclysmic variables, their origin is somewhat 
doubtful. The radiative torques speculated to be responsible for 
the better established warps in X-ray binaries definitely do 
not generally operate in CVs, leaving no clear mechanism 
that is known to be effective in creating a tilted disc. 
We note that of those
systems that do show negative superhumps, TV Col is a known
intermediate polar (Hellier 1993), and V603 AqI is a suspected one
(Schwarzenberg-Czerny, Udalski \& Monier 1992).
That then implies that the dynamics of the inner
disc are significantly influenced by the magnetic field.
The details of the r\^ole a magnetic field might play in creating a
warped disc are not clear, however we note that Agapitou, Papaloizou
\& Terquem (1997) have shown that {\em strongly} magnetised thin discs are
unstable to bending modes.
Further investigation of the
correlation between negative superhumps and weakly magnetic CVs would be
very interesting.

\section*{Acknowledgments}
The computations were completed using the {\small SGI} Power Challenge
computer at The Australian National University Supercomputer Facility.

\bsp
\end{document}